\documentclass[aps,prb,reprint,nofootinbib,amsmath,amssymb,superscriptaddress]{revtex4-2}

\usepackage{amsmath,amssymb,amsthm,bm,physics,mathtools,graphicx,booktabs,enumitem,hyperref,array,float}

\usepackage[margin=1.8cm]{geometry}

\newcommand{\Geff}{\Gamma_{\rm eff}}
\newcommand{\ave}[1]{\langle #1\rangle}
\newcommand{\bk}[3]{\langle #1|#2|#3\rangle}

\begin{document}

\title{Exact solution of the dynamic pseudogap model:\\ Fock-space formalism 
for Gaussian fluctuations}

\author{Michael V. Sadovskii}

\affiliation{Instutute for Electrophysics, Russian Academy of Science,
Ural Branch, Amundsen str. 106, Ekaterinburg 620016, Russia}

\begin{abstract}
We present an exact solution of the dynamic finite-$Q$ pseudogap model, 
in which electrons scatter between two nested Fermi-surface patches off 
a classical Gaussian order-parameter field with characteristic frequency
$\omega_0$ and damping $\gamma$.
The disorder-averaged Green's function is shown to be a single matrix 
element of the resolvent of a sparse non-Hermitian matrix acting on a 
Fock space of two (commensurate fluctuations) or four (incommensurate 
fluctuations) bosonic modes with complex energies. The non-commutativity 
of the pseudospin vertices at different times, which prevented an exact 
treatment so far, is retained in full. We derive the representation, 
its symmetries, and the exact limits it contains: the static continued 
fractions, the closed-form commensurate solution at the Fermi surface, 
and the second-order self-energy. 
A numerically exact implementation gives spectral functions and densities of 
states for all values of the fluctuation frequency, damping and electron energy. 
We find that the sideband comb of the spectral function stays close to multiples 
of $\omega_0$ and that quasiparticle damping is governed by an energy-dependent 
decoherence rate. 
At resonance of a quasiparticle energy with $\omega_0/2$,  we obtain
the static fuctuating-gap in a rotating frame, which leads to resonant 
splitting of the quasiparticle peak.

\end{abstract}

\maketitle

\section{Introduction}

Pseudogap formation by strong short-range order fluctuations is described, 
in a number of cases, by exactly solvable models in which electrons scatter 
off a classical Gaussian random field with a finite ordering vector 
$\mathbf Q$ and a finite correlation length 
\cite{Sad74a,Sad74b,Sad79,SadDiag,SPS98,SPS99,KS99}. 
For a \emph{static} field the full diagrammatic series is summed by a 
continued-fraction recursion. The extension to fluctuations with a 
finite correlation time is an open problem, relevant whenever the 
characteristic frequency of the order-parameter fluctuations is comparable 
to electronic energies.

For a spatially uniform field ($Q=0$) the dynamic problem is exactly 
solvable \cite{KK09,EK22,KS24}. For $Q\neq0$ every scattering event transfers 
the electron between the nested branches $\xi$ and $-\xi$, and the Nambu 
Hamiltonian $\hat H(t)=\xi\sigma_3+\Delta(t)\sigma_-+\Delta^*(t)\sigma_+$ 
does not commute with itself at different times. 
A heuristic double-series description based on an Abelian (commuting) 
projection of the pseudospin algebra was proposed in Ref.~\cite{KS26}, where
the detailed discussion of the physics of this model can also be found. 
A numerically exact Monte Carlo solution was then used in Ref.~\cite{S26} 
to delimit its range of validity.

In this paper we show that the problem is {\em exactly} solvable in a different 
sense: the averaged Green's function of the model with a correlator 
$\Delta^2e^{-\gamma|t|}\cos\omega_0t$ is an exact resolvent matrix element 
on a Fock space of bosonic modes with complex energies. 
Nothing is dropped: the time ordering of the pseudospin vertices is treated 
exactly, and the cost of doing so appears only as the dimension of a 
sparse matrix. The purpose of this paper is to present the formalism, 
the exact analytic results that follow from it, and the physical picture 
that the exact numerical solution gives. 
Comparisons with approximate treatments are kept to a minimum and are made 
only where they illuminate the structure of the exact solution.


\section{Model}\label{sec:model}

Near nested portions of the Fermi surface, with 
$\xi_{\mathbf k+\mathbf Q}\simeq-\xi_{\mathbf k}\equiv-\xi$, the electron 
is described in the Nambu basis $(c_{\mathbf k},c_{\mathbf k+\mathbf Q})$ by
\begin{equation}
\hat H(t)=\xi\sigma_3+\Delta_1(t)\sigma_1+\Delta_2(t)\sigma_2=\xi\sigma_3
+\Delta\sigma_-+\Delta^*\sigma_+,
\label{eq:H}
\end{equation}
with $\sigma_\pm=(\sigma_1\pm i\sigma_2)/2$ and 
$\Delta=\Delta_1+i\Delta_2$. 
The fluctuating field is a zero-mean stationary Gaussian process,
\begin{equation}
\ave{\Delta(t)\Delta^*(0)}=\Delta^2e^{-\gamma|t|}\cos\omega_0t,
\label{eq:corr}
\end{equation}
where $\Delta$ is the fluctuation amplitude, $\gamma^{-1}$ the correlation 
time and $\omega_0$ the characteristic oscillation frequency. 
\emph{Commensurate} fluctuations are described by a real field ($\Delta_2=0$). 
For \emph{incommensurate} fluctuations the field is complex. 
The quantity of interest is the disorder-averaged 
retarded Green's function
\begin{equation}
G^R(t)=-i\theta(t)\,\ave{U_{11}(t)},\quad U(t)
=T\exp\Big[-i\!\int_0^t\!\hat H(t')dt'\Big],
\label{eq:GRt}
\end{equation}
and the spectral density 
$A(\omega,\xi)=-\pi^{-1}\mathrm{Im}\,G^R(\omega+i\eta,\xi)$, 
where $\eta$ is a small broadening. 
The density of states is $N(\omega)=\int d\xi\,A(\omega,\xi)$ for a band of 
infinite width and exact nesting. Energies are in units of $\Delta$ unless 
stated otherwise.

The difficulty is that $[\hat H(t_1),\hat H(t_2)]\neq0$ in 
general \cite{KS26}: the commutator vanishes only for $\xi=0$ and a real field. 
We show below that this does not prevent an exact evaluation 
of the \emph{average}.

\section{Exact Fock-space representation}\label{sec:formalism}
\subsection{Gaussian field as a bosonic bath}

Consider a circular complex Ornstein--Uhlenbeck (OU) process $z(t)$ 
with $\ave{z(t)z^*(0)}=s\,e^{-\kappa t}$ for $t>0$ ($\mathrm{Re}\,\kappa>0$), 
$\ave{zz}=0$, and (by stationarity and conjugation) 
$\ave{z(t)z^*(0)}=s\,e^{-\kappa^*|t|}$ for $t<0$. Introduce two bosonic 
modes $a,b$ with the ``bath Hamiltonian''
\begin{equation}
H_B=\varepsilon_a a^\dagger a+\varepsilon_b b^\dagger b,\quad \varepsilon_a=
-i\kappa,\quad \varepsilon_b=-i\kappa^*,
\label{eq:eps}
\end{equation}
and the linear forms $z=\sqrt s(a+b^\dagger)$, $z^*=\sqrt s(a^\dagger+b)$. 
For $\kappa=\gamma+i\omega_0$,
\begin{equation}
\varepsilon_a=\omega_0-i\gamma,\qquad \varepsilon_b=-\omega_0-i\gamma,
\end{equation}
both with non-positive imaginary part. In the interaction picture 
$a(t)=ae^{-i\varepsilon_at}$, $a^\dagger(t)=a^\dagger e^{i\varepsilon_at}$, 
and similarly for $b$. With $\ket0$ the bath vacuum and for $t_1>t_2$,
\begin{align}
\bk0{z(t_1)z^*(t_2)}0&=s\,e^{-\kappa(t_1-t_2)},\\
\bk0{z^*(t_1)z(t_2)}0&=s\,e^{-\kappa^*(t_1-t_2)},\\
\bk0{z(t_1)z(t_2)}0&=0,
\end{align}
which are exactly the two-point functions of the process $z$ for both time
orderings.

Now consider $V(t)=\sum_\alpha\Phi_\alpha(t)M_\alpha$, with $M_\alpha$ 
fixed matrices in pseudospin space and $\Phi_\alpha$ either the classical 
Gaussian field or the corresponding linear bath operator, with the same 
two-point function 
$C_{\alpha\beta}(t,t')=\bk0{T\Phi_\alpha(t)\Phi_\beta(t')}0$. 
Then it is easily shown that
\begin{equation}
\Big\langle T e^{-i\int_0^t(\hat H_{\rm el}+V)}\Big\rangle_{\rm field}=
\bk0{Te^{-i\int_0^t(\hat H_{\rm el}+V)}}0 .
\end{equation}
To prove, just expand both sides in powers of $V$. 
Each term is an integral over ordered times of a product of $c$-number 
matrices in time order times an $n$-point average of $\Phi$'s. 
For zero-mean Gaussian variables the average of an odd number vanishes 
and for an even number it is the sum over pairings of two-point functions. 
For linear forms in $a,a^\dagger,b,b^\dagger$ the vacuum satisfies Wick's 
theorem, again giving the sum over pairings, with the order inside each pair 
given by the time ordering. The matrices commute with the bath operators. 
Hence both sides have the same pairings, pair weights and matrix products in 
time order. Wick's theorem for linear forms does not require Hermiticity, 
so complex $\varepsilon$ is allowed.
No commutator of the matrices $M_\alpha$ is dropped.

\subsection{Reduction to a resolvent}

With $\hat H_{\rm eff}=\hat H_{\rm el}+H_B+V$ (Schr\"odinger picture) and 
$\bra0e^{iH_Bt}=\bra0$,
\begin{equation}
\ave{U(t)}=\bk0{e^{-i\hat H_{\rm eff}t}}0 .
\end{equation}
For $\ket{\psi_0}=\ket\uparrow\otimes\ket0$ the retarded function follows 
by Laplace transform,
\begin{equation}
\boxed{G^R(\omega,\xi)=
\bk{\psi_0}{(\omega-\hat H_{\rm eff})^{-1}}{\psi_0},
\quad \mathrm{Im}\,\omega>0.}
\label{eq:GR}
\end{equation}
The integral converges because 
$\mathrm{Im}\,\hat H_{\rm eff}=-\gamma\hat N\le0$, with $\hat N$ the total 
boson number.

\subsection{Commensurate fluctuations}

Let $\Delta_1=\mathrm{Re}\,z$ with $\kappa=\gamma+i\omega_0$ and 
$s=2\Delta^2$. 
Then $\ave{\Delta_1(t)\Delta_1(0)}=
\tfrac14s(e^{-\kappa|t|}+e^{-\kappa^*|t|})=
\Delta^2e^{-\gamma|t|}\cos\omega_0t$, and
\begin{equation}
\hat H_{\rm eff}^{\rm c}
=\xi\sigma_3+\varepsilon_an_a+\varepsilon_bn_b+
\frac{\Delta}{\sqrt2}\sigma_1(a+a^\dagger+b+b^\dagger).
\label{eq:Hc}
\end{equation}
The operator $\Pi=\sigma_3(-1)^{n_a+n_b}$ commutes with 
$\hat H_{\rm eff}^{\rm c}$ and equals $+1$ on $\ket{\psi_0}$.

\subsection{Incommensurate fluctuations}

Let $\Delta=z_1+z_2$ with $z_1$ ($\kappa_1=\gamma+i\omega_0$) and 
$z_2$ ($\kappa_2=\gamma-i\omega_0$) independent, each with $s=\Delta^2/2$. 
Then $\ave{\Delta\Delta^*}=
\tfrac{\Delta^2}2(e^{-\kappa_1|t|}+e^{-\kappa_2|t|})$ and 
$\ave{\Delta\Delta}=0$. With four modes $a_{1,2},b_{1,2}$,
\begin{align}
\hat H_{\rm eff}^{\rm i}&=\xi\sigma_3+\sum_j\varepsilon_jn_j\nonumber\\
&\quad+\frac{\Delta}{\sqrt2}\Big[(a_1+b_1^\dagger+a_2+b_2^\dagger)
\sigma_-+\text{h.c.}\Big],
\label{eq:Hi}
\end{align}
with $\varepsilon_{a_1}=\varepsilon_{b_2}=\omega_0-i\gamma$ and 
$\varepsilon_{a_2}=\varepsilon_{b_1}=-\omega_0-i\gamma$. 
The charge $C=\tfrac12\sigma_3-n_a+n_b$, with $n_a=n_{a_1}+n_{a_2}$ and 
$n_b=n_{b_1}+n_{b_2}$, is conserved: both $\sigma_-a$ and $\sigma_-b^\dagger$ 
lower $\sigma_3/2$ by one and raise $-n_a+n_b$ by one. 
The initial state has $C=\tfrac12$.

\begin{figure*}
\includegraphics[width=0.92\textwidth]{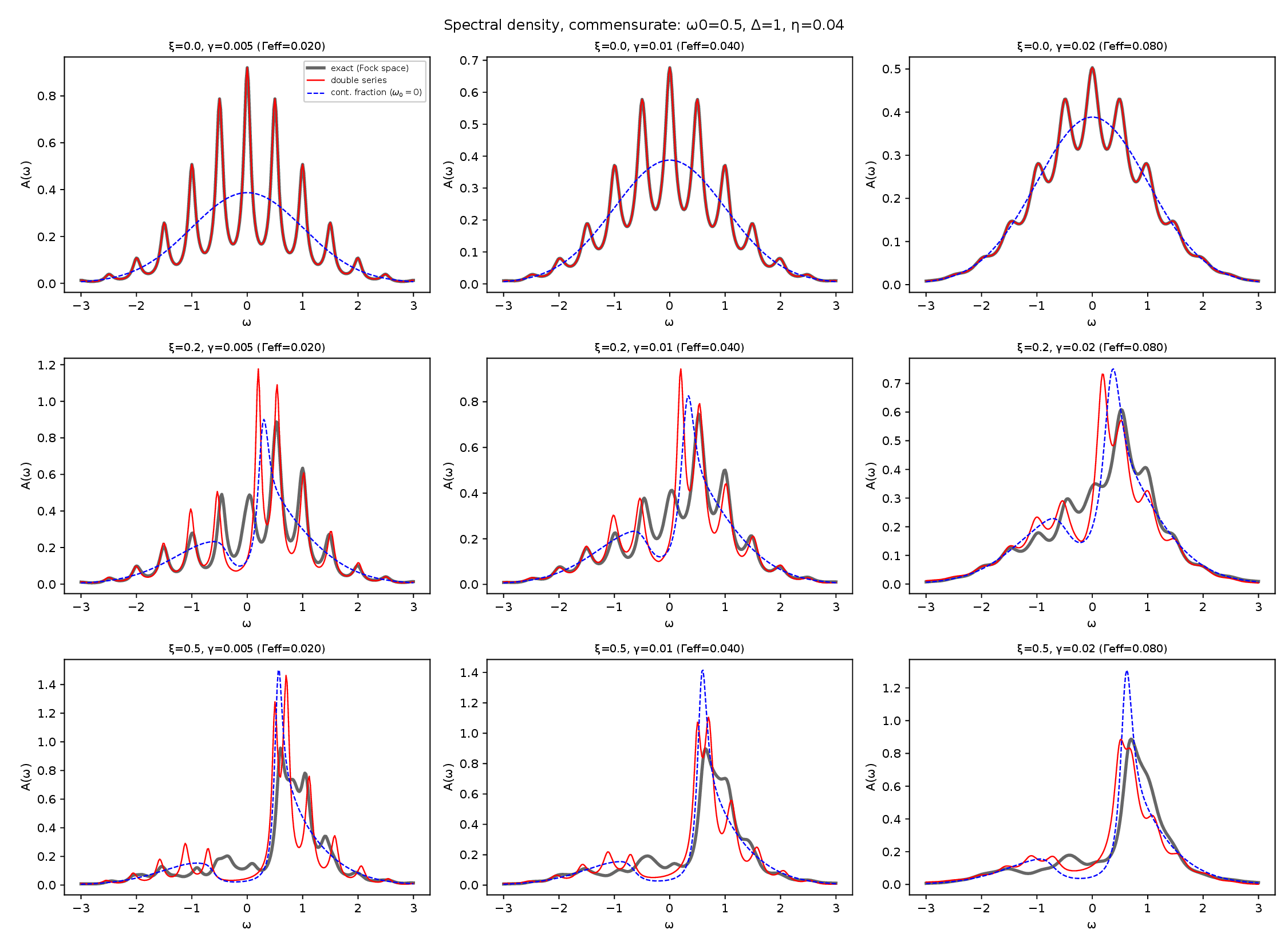}
\caption{Exact spectral density for commensurate fluctuations 
($\omega_0=0.5$, $\Delta=1$, $\eta=0.04$) at $\xi=0,0.2,0.5$ (rows) 
and $\gamma=0.005,0.01,0.02$ (columns). 
Black: exact Fock-space solution. Red and blue are the approximate double 
series and static continued fraction results \cite{KS26}, shown here and below 
for reference only.}
\label{fig:sc}
\end{figure*}

\section{Exact analytic results}\label{sec:exact}
\subsection{Static limit: continued fractions}

For $\omega_0=0$ the two incommensurate modes have the same $\kappa=\gamma$, 
so $z_1+z_2$ is a single circular OU process of variance $\Delta^2$: 
one pair $(a,b)$ with $\varepsilon_a=\varepsilon_b=-i\gamma$ and coupling
$\Delta(a+b^\dagger)\sigma_-+\text{h.c.}$ The states reached from
$\ket{\psi_0}$ are $\ket{\uparrow;n,n}$ and $\ket{\downarrow;n,n+1}$ and form a
chain: site $k$ has $k$ quanta, energy $(-1)^k\xi-ik\gamma$, and the hopping
between sites $k-1$ and $k$ is $\Delta\sqrt{\lceil k/2\rceil}$. Hence
\begin{equation}
G=\cfrac{1}{\omega-\xi-\cfrac{v(1)\Delta^2}{\omega+\xi+i\gamma-
\cfrac{v(2)\Delta^2}{\omega-\xi+2i\gamma-\cdots}}},
\end{equation}
with $v(k)=\lceil k/2\rceil$. For commensurate fluctuations at 
$\omega_0=0$, the modes combine into $c=(a+b)/\sqrt2$ plus a decoupled mode; 
the coupling becomes $\Delta\sigma_1(c+c^\dagger)$ and the chain has hopping 
$\Delta\sqrt k$, i.e.\ $v(k)=k$. These are the exact static recursions obtained 
earlier by diagram summation \cite{Sad79,SadDiag}; here they follow without 
diagram combinatorics, and the factors $v(k)$ appear as boson-number matrix 
elements.

\subsection{Commensurate fluctuations at the Fermi surface}

At $\xi=0$ the Hamiltonian $\Delta_1(t)\sigma_1$ commutes at all times, 
$U_{11}=\cos X$ with $X=\int_0^t\Delta_1$ Gaussian, and 
$\ave{U_{11}}=e^{-K(t)}$ with \cite{KS26}
\begin{equation}
K(t)=\sum_\pm\frac{\Delta^2}{2(\gamma\pm i\omega_0)^2}
\Big[(\gamma\pm i\omega_0)t-1+e^{-(\gamma\pm i\omega_0)t}\Big].
\end{equation}
The linear part of $K$ gives the long-time decay rate 
$\Geff=\gamma\Delta^2/(\omega_0^2+\gamma^2)$ of the amplitude, 
i.e.\ a half width at half maximum $\Geff$ of the corresponding spectral 
features. The representation \eqref{eq:GR} reproduces this solution numerically 
(Sec.~\ref{sec:num}).

\subsection{Second order and an energy-dependent decoherence rate}
\label{sec:born}

At second order the intermediate states are the one-boson states 
$\ket{\downarrow;a}$, $\ket{\downarrow;b}$ with amplitude $\Delta/\sqrt2$ and 
energy $-\xi+\varepsilon$. The self-energy is
\begin{equation}
\Sigma^{(2)}(\omega,\xi)=\frac{\Delta^2}{2}
\Big[\frac{1}{\omega+\xi-\varepsilon_a}+\frac1{\omega+\xi-\varepsilon_b}\Big].
\label{eq:Sigma2}
\end{equation}
The incommensurate model gives the same expression because only 
$b_1^\dagger$ and $b_2^\dagger$ act on $\ket{\uparrow;0}$ through $\sigma_-$ 
and $\{\varepsilon_{b_1},\varepsilon_{b_2}\}=\{\varepsilon_b,\varepsilon_a\}$. 
Evaluating at the pole, $\omega\simeq\xi$, gives a quasiparticle energy 
$\xi+\mathrm{Re}\,\Sigma^{(2)}$ and a half width
\begin{equation}
\Gamma(\xi)=\frac{\Delta^2\gamma}{2}
\Big[\frac1{(2\xi-\omega_0)^2+\gamma^2}+\frac1{(2\xi+\omega_0)^2
+\gamma^2}\Big].
\label{eq:Gxi}
\end{equation}
Two features follow. First, $\Gamma(0)=\Geff$: the decoherence rate of the 
earlier theory is the Fermi-surface value of a rate that depends on $\xi$. 
Second, for $\gamma\to0$,
\begin{equation}
\Gamma(\xi)\to\frac{\pi\Delta^2}{4}\big[\delta(2\xi-\omega_0)
+\delta(2\xi+\omega_0)\big],
\end{equation}
a golden-rule rate for converting a quasiparticle of energy $\xi$ into 
the $-\xi$ branch by exchanging a quantum $\pm\omega_0$ with the field; 
it is resonant at $2|\xi|=\omega_0$. The satellites of the Born spectrum 
sit at $\omega=-\xi\pm\omega_0$ with weights $\Delta^2/[2(\omega_0\mp2\xi)^2]$. 
At large $\gamma$, Eqs.~\eqref{eq:Sigma2}--\eqref{eq:Gxi} are expected to 
become accurate; we test this in Sec.~\ref{sec:results}.

\section{Numerical implementation and tests}\label{sec:num}
\subsection{Algorithm}
The matrices \eqref{eq:Hc} and \eqref{eq:Hi} are sparse and complex symmetric. 
We truncate the boson numbers and, for each frequency, solve 
$(\omega+i\eta-\hat H_{\rm eff})x=e_{\psi_0}$ by sparse LU factorization, 
so that $A(\omega,\xi)=-\pi^{-1}\mathrm{Im}\,x_{\psi_0}$. 
The broadening enters as $\omega+i\eta$, i.e.\ it is equivalent to a window 
$e^{-\eta t}$ on the time-domain average. The density of states is obtained by 
integrating over $\xi$ with the trapezoid rule on a uniform grid of step 
$\lesssim\eta$ and a Lorentzian tail correction. More details can be found in
Appendix A.

\subsection{Tests}
The following checks were performed:
\begin{itemize}
\item \textbf{Closed form at $\xi=0$.} 
For commensurate fluctuations with $\omega_0=0.5$, $\gamma=0.02$, $\eta=0.02$: 
$A(\omega=0)=0.54475$ from the closed form and from the Fock-space solver 
(agreement of the spectra at $\omega=0,\pm0.5$ to $3\times10^{-6}$).
\item \textbf{Static limits.} For $\omega_0=0$, $\xi=0.2$, $\gamma=0.1$, 
both models agree with the continued fractions to better than $10^{-7}$ at 
the frequencies tested.
\item \textbf{Sum rule.} 
For $\xi=0.2$, $\omega_0=0.5$, $\gamma=0.02$, $\eta=0.1$ (grid step 
$0.05$, range $[-30,30]$), $\int A\,d\omega$ plus the Lorentzian tail equals 
$1$ to $3\times10^{-6}$ for both models.
\item \textbf{Free electrons.} For $\Delta\to0$ the density of states 
approaches $1$ to $10^{-5}$.
\end{itemize}
\textbf{Truncation.} The weight of a bath state decays with the number of 
quanta, but the cutoff needed grows with the Poisson-type parameter 
$\lambda=\Delta^2/[2(\omega_0^2+\gamma^2)]$ \cite{KS26}. 
For commensurate $\xi=0.2$, $\omega_0=0.5$, $\gamma=0.02$ the maximum change 
of $A$ relative to a cutoff-70 reference is $1.1\times10^{-2}$, 
$4.0\times10^{-3}$, $6\times10^{-4}$ and $3\times10^{-5}$ for cutoffs 
$26$, $31$, $38$ and $50$. For incommensurate $\xi=0$, cutoffs $14$ and 
$20$ differ by less than $10^{-3}$. A systematic cutoff study for all 
parameters shown below was not performed; the incommensurate results at the 
smallest $\gamma$ should be regarded as less converged.

\begin{figure*}
\includegraphics[width=0.92\textwidth]{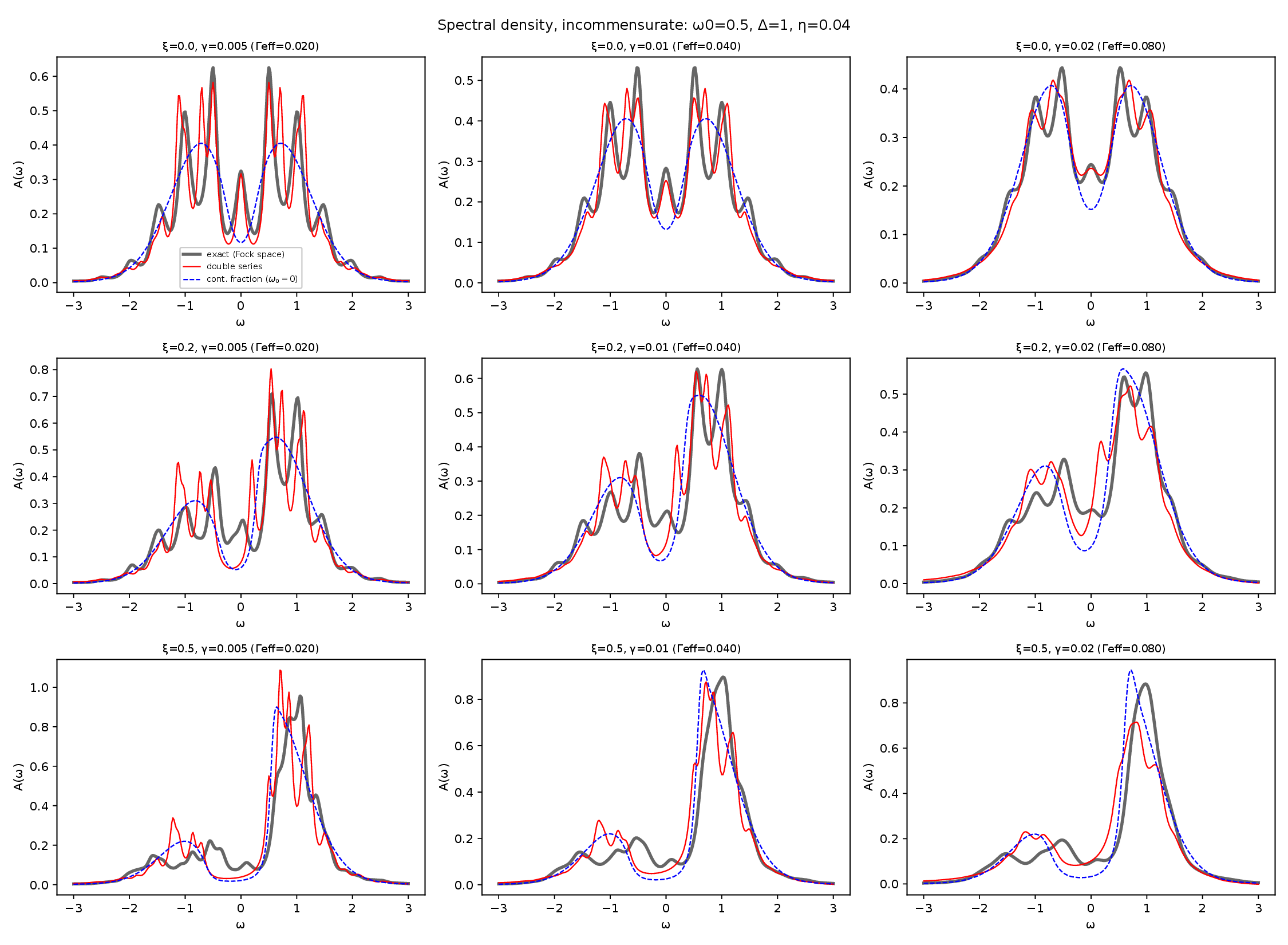}
\caption{As Fig.~\ref{fig:sc} for incommensurate fluctuations 
(cutoff $n_{\max}=16$).}\label{fig:si}
\end{figure*}

\section{Results}\label{sec:results}

\subsection{Spectral functions}

Figures~\ref{fig:sc} and \ref{fig:si} show exact spectral functions for the 
two types of fluctuations. At the Fermi surface and for commensurate 
fluctuations the spectrum is a comb of sidebands at $\omega=n\omega_0$ 
whose envelope is the static spectral function; the comb is washed out 
as $\gamma$ (and $\Geff=0.02,0.04,0.08$) grows. This is the exactly solvable 
case of Sec.~\ref{sec:exact}.

Away from the Fermi surface the exact spectrum contains a quasiparticle-like 
main peak near $\omega\approx\xi$ dressed by sidebands. A notable property of 
the exact solution at small $\xi$ is that the sideband comb stays close to 
multiples of $\omega_0$ rather than following 
$\pm\sqrt{\xi^2+n^2\omega_0^2}$ as in Ref. \cite{KS26}. 
At $\gamma=0.005$ and $\xi=0.2$, the peaks of $A$ are at 
$\omega=-0.98,-0.46,0.04,0.52,1.0,1.48$ (commensurate) and 
$-0.98,-0.46,0.04,0.56,1.02,1.44$ (incommensurate), 
i.e.\ within about $0.06$ of $n\omega_0$. At $\xi=0.5$ the comb is distorted 
by the main peak (commensurate peaks at $-0.36,0.08,0.6,1.04,1.4$) and the 
sidebands of the two branches no longer line up with a single comb. 
The incommensurate spectra show a weaker modulation than the commensurate ones, 
presumably because each sideband is split by the two independent circular 
components of the field; we have not tested this explanation.


\begin{figure}
\includegraphics[width=0.95\columnwidth]{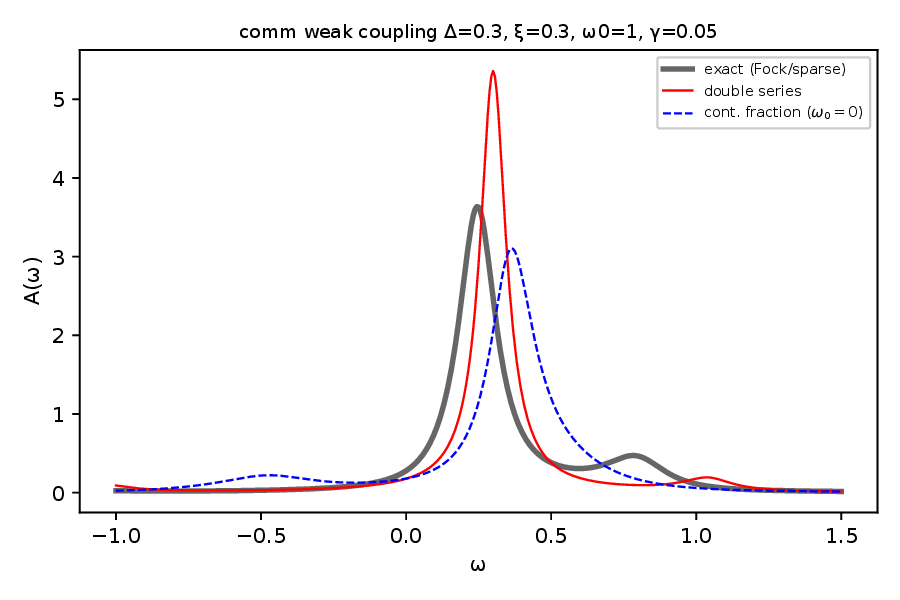}
\caption{Weak coupling ($\Delta=0.3$, $\xi=0.3$, $\omega_0=1$, 
$\gamma=0.05$, $\eta=0.05$, commensurate). The exact solution has a main peak 
of height $3.64$ at $\omega=0.245$ and a satellite at $\omega=0.78$, close to 
the Born position $-\xi+\omega_0=0.7$ shifted by the self-energy.}
\label{fig:weak}
\end{figure}
For $\Delta\ll\omega_0$ the exact solution agrees with the perturbative picture 
of Sec.~\ref{sec:born} (Fig.~\ref{fig:weak}): one main peak and a satellite at 
$\omega\simeq-\xi+\omega_0$, shifted to $0.78$ by $\mathrm{Re}\,\Sigma^{(2)}$ 
(Ref.~\cite{S26} quotes $0.79$ from second-order perturbation theory).

\begin{figure*}
\includegraphics[width=0.92\textwidth]{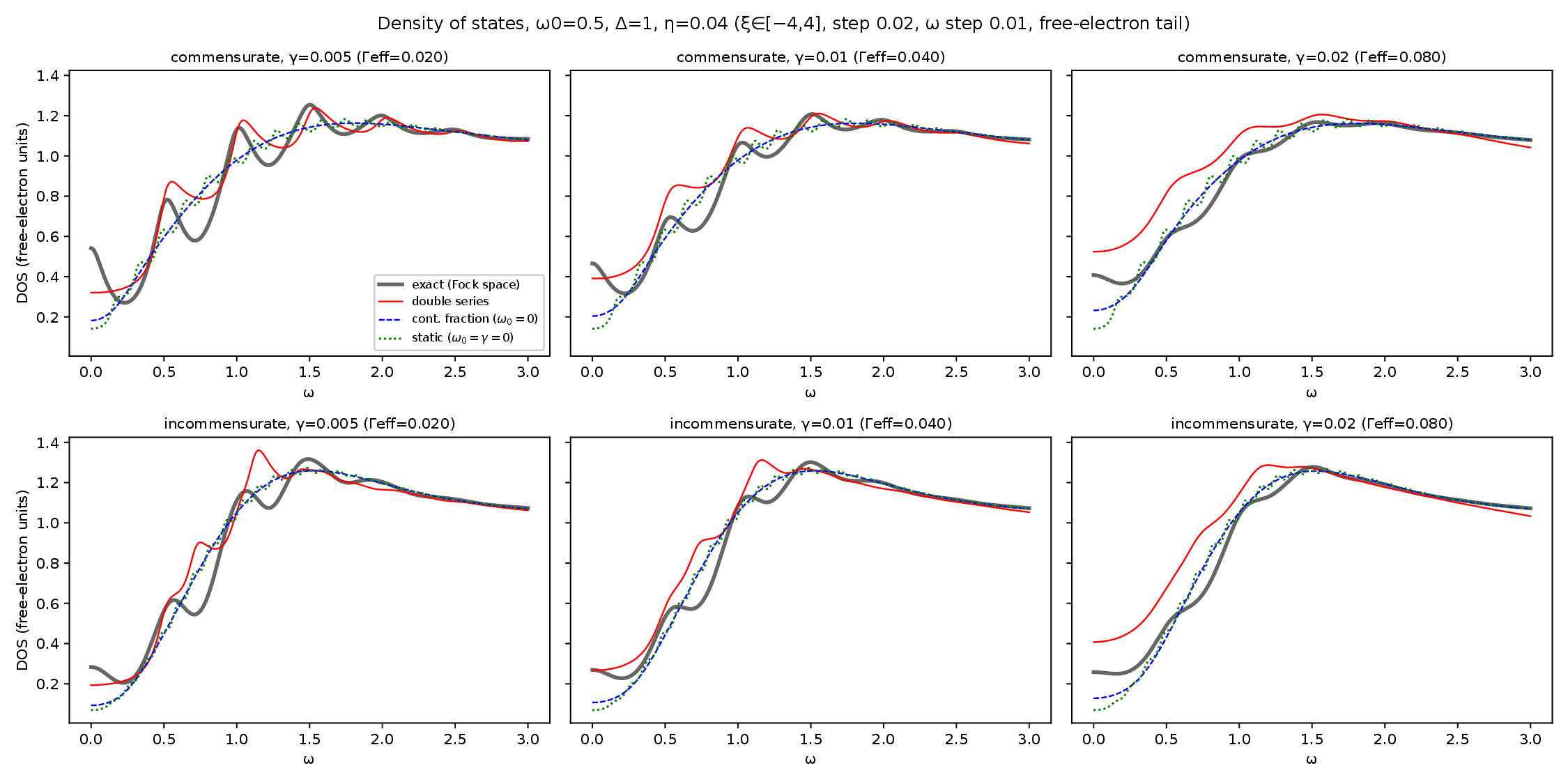}
\caption{Exact density of states (units of the free-electron value) for 
commensurate (top) and incommensurate (bottom) fluctuations 
at $\omega_0=0.5$, $\Delta=1$, $\eta=0.04$, and $\gamma=0.005,0.01,0.02$. 
The $\xi$ integral runs over $[-4,4]$ with step $0.02$ and a 
free-electron tail. The frequency step is $0.01$. 
Cutoffs are $n_{\max}=80$ (commensurate) and $n_{\max}=30$
(incommensurate).}\label{fig:dos}
\end{figure*}

\subsection{Density of states}

Figure~\ref{fig:dos} shows exact densities of states. For commensurate 
fluctuations the pseudogap is modulated by structures at multiples of 
$\omega_0$ (near $\omega\simeq0.5,1,1.5$) that are suppressed with growing 
$\gamma$, and the value at $\omega=0$ is $0.54$, $0.47$, $0.41$ for 
$\gamma=0.005,0.01,0.02$: the pseudogap is partly filled even at the 
smallest $\gamma$; this is presumably due to the sidebands of the spectral 
function at $\xi\neq0$ reaching $\omega=0$, which we have not verified 
separately. For incommensurate fluctuations the modulation is weaker, 
and $N(0)=0.28$, $0.27$, $0.26$ for the same values of $\gamma$. 

All numerical results presented here are in complete agreement with the
results of direct Monte Carlo simulations of Ref. \cite{S26}.

\section{Resonance at $\xi=\omega_0/2$}\label{sec:res}

The $\xi$-dependent rate $\Gamma(\xi)$ of Eq.~\eqref{eq:Gxi} has maxima at
$2|\xi|=\omega_0$. In the Fock-space formulation this resonance has a simple
origin, can be treated in closed form, and changes the spectral function
qualitatively.

\subsection{Degeneracy in the Fock space}

In $\hat H_{\rm eff}$ the initial state $\ket{\uparrow;{\rm vac}}$ has
energy $\xi$. The states reached by one transition have a pseudospin flipped
and one boson added, with energy $-\xi+\varepsilon$. For the incommensurate
model the relevant boson is $b_2$ with $\varepsilon_{b_2}=\omega_0-i\gamma$,
for the commensurate model it is $a$ with $\varepsilon_a=\omega_0-i\gamma$.
The two levels are degenerate (up to the damping) when
\begin{equation}
2\xi=\omega_0 .
\label{eq:rescond}
\end{equation}
A selection rule restricts the resonances. In the commensurate model
$\Pi=\sigma_3(-1)^{n_a+n_b}$ is conserved, so states with $\sigma_3=-1$ have
an odd number of quanta, and the real part of their bath energy,
$(n_a-n_b)\omega_0$, is an odd multiple of $\omega_0$.
In the incommensurate model the sector $C=\tfrac12$ has $n_b=n_a+1$ for
$\sigma_3=-1$, so the number of quanta is odd and the real part of the bath
energy is again an odd multiple of $\omega_0$. States with $\sigma_3=+1$ have
even multiples. Degeneracy with $\ket{\uparrow;{\rm vac}}$ therefore requires
\begin{equation}
2\xi=(2m+1)\omega_0,\qquad m=0,1,2,\dots
\end{equation}
The case $m=0$ is a first-order resonance. Values $m\ge1$ require three or
more quanta. In particular $\xi=\omega_0$ is \emph{not} a resonance, although
the sidebands are strongly distorted there in Figs.~\ref{fig:sc} and
\ref{fig:si}.

\subsection{Resonant mapping onto the static fluctuating-gap model}

Near $2\xi=\omega_0$ only the resonant terms of $\hat H_{\rm eff}$ are kept
(rotating-wave approximation). For the incommensurate model these are the terms
containing the mode pair $(a_2,b_2)$, i.e.\ the single circular component
$z_2$ of variance $\Delta^2/2$:
\begin{equation}
\hat H_{\rm res}=\xi\sigma_3+\varepsilon_{a_2}n_{a_2}+
\varepsilon_{b_2}n_{b_2}+\frac{\Delta}{\sqrt2}
\big[(a_2+b_2^\dagger)\sigma_-+\text{h.c.}\big].
\end{equation}
For the commensurate model the real field contains $z$ and $z^*$,
$\Delta_1=\tfrac12(z+z^*)$ with $s=2\Delta^2$, and the resonant part is
$\tfrac12z^*\sigma_-+\text{h.c.}$ A resonant component of variance
$s/4=\Delta^2/2$ appears, so the resonant problem is the same.
The states $\ket{\uparrow;n,n}$ and $\ket{\downarrow;n,n+1}$ in the two modes
form a chain $k=0,1,\dots$ with energies and hoppings
\begin{eqnarray}
&E_{2n}=\xi-2in\gamma,\quad E_{2n+1}=-\xi+\omega_0-i(2n+1)\gamma,
\,\nonumber\\ 
&t_k=\frac{\Delta}{\sqrt2}\sqrt{\lceil k/2\rceil},
\end{eqnarray}
so that
\begin{equation}
G_{\rm res}(\omega,\xi)=\cfrac{1}{\omega-\xi-\cfrac{v(1)\Delta^2/2}
{\omega+\xi-\omega_0+i\gamma-\cfrac{v(2)\Delta^2/2}
{\omega-\xi+2i\gamma-\cdots}}}
\label{eq:Gres}
\end{equation}
with $v(k)=\lceil k/2\rceil$. With $\Omega=\omega-\omega_0/2$ and
$\tilde\xi=\xi-\omega_0/2$ the denominators are $\Omega-\tilde\xi$,
$\Omega+\tilde\xi+i\gamma$, $\Omega-\tilde\xi+2i\gamma,\dots$, which is
exactly the static incommensurate continued fraction of Sec.~\ref{sec:exact}
with
\begin{equation}
G_{\rm res}(\omega,\xi)=G^{\rm static}_{\rm inc}\big(\omega-\tfrac{\omega_0}2,\
\xi-\tfrac{\omega_0}2;\ \Delta\to\Delta/\sqrt2,\ \gamma\big).
\label{eq:map}
\end{equation}
Near the resonance, the dynamic problem is therefore a \emph{static}
fluctuating-gap problem in a rotating frame: its ``Fermi surface'' sits at
$\xi=\omega_0/2$ and its pseudogap at $\omega=\omega_0/2$, with the amplitude
reduced by $\sqrt2$. For $\gamma\to0$ and $\tilde\xi=0$ the average over the
Rayleigh distributed amplitude gives, with $x=\omega-\omega_0/2$ and
$\sigma^2=\Delta^2/2$,
\begin{equation}
A(\omega)=\frac{|x|}{\sigma^2}\,e^{-x^2/\sigma^2},
\label{eq:Rayleigh}
\end{equation}
a doublet with maxima at $x=\pm\Delta/2$ and a zero at $x=0$: the resonance
\emph{splits} the quasiparticle peak and produces a pseudogap at its centre.
By the symmetry $A(\omega,\xi)=A(-\omega,-\xi)$, verified numerically to
$10^{-14}$ for both models, there is a mirror feature at $\xi=-\omega_0/2$,
$\omega=-\omega_0/2$.

\subsection{Counter-rotating corrections}

The terms dropped in Eq.~\eqref{eq:Gres} shift the levels.
For the incommensurate model the leading effect is the second-order shift of
$\ket{\uparrow;{\rm vac}}$ by the non-resonant intermediate state
$\ket{\downarrow;b_1}$ with energy $-\xi-\omega_0$:
\begin{equation}
\delta=\frac{\Delta^2/2}{\omega+\xi+\omega_0}
\Big|_{\omega=\xi=\omega_0/2}=\frac{\Delta^2}{4\omega_0},
\end{equation}
which enters as $\tilde\xi\to\tilde\xi+\Delta^2/(4\omega_0)$ in
Eq.~\eqref{eq:map} (Bloch--Siegert-type shift).
A three-parameter fit of Eq.~\eqref{eq:map} (centre shift, detuning shift,
amplitude scale) to the exact spectrum at $\xi=\omega_0/2$ gives incommensurate
detuning shifts $0.0049$, $0.0185$, $0.0381$ for $\Delta=0.1,0.2,0.3$
(analytic $\Delta^2/4\omega_0=0.005,0.02,0.045$), amplitude scales
$0.999,0.992,0.973$, and residual maximum deviations of $0.7\%$, $2.4\%$,
$5.7\%$ of the peak height. For commensurate fluctuations the fitted detuning
shifts are smaller, $0.0029$, $0.0104$, $0.0222$, together with centre shifts
$0.0019,0.0084,0.0195$ and amplitude scales $0.988,0.953,0.886$; the residuals
are $1.0\%$, $3.7\%$, $10\%$. For the commensurate model the simple analytic
shift above therefore does not suffice.

\subsection{Numerical results}

\begin{figure*}
\includegraphics[width=0.9\textwidth]{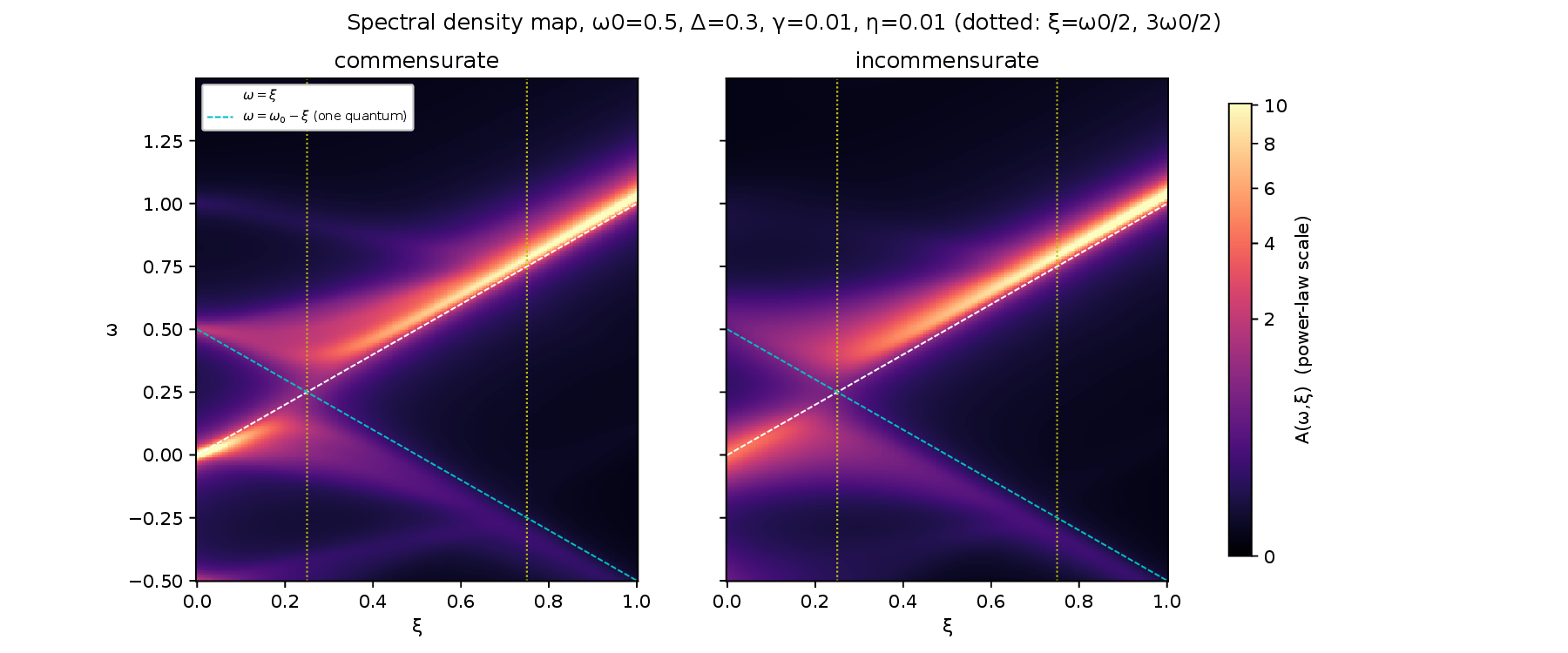}
\caption{Exact spectral density $A(\omega,\xi)$ for $\omega_0=0.5$,
$\Delta=0.3$, $\gamma=0.01$, $\eta=0.01$ (power-law colour scale).
White dashed: $\omega=\xi$; cyan dashed: one-quantum satellite
$\omega=\omega_0-\xi$; yellow dotted: $\xi=\omega_0/2,3\omega_0/2$.
The two branches repel each other at $\xi=\omega_0/2$.}
\label{fig:resmap}
\end{figure*}
Figure~\ref{fig:resmap} shows $A(\omega,\xi)$ across the resonance.
The main branch $\omega\simeq\xi$ and the one-quantum satellite branch
$\omega=\omega_0-\xi$ cross at $\xi=\omega_0/2$, and the spectral weight
follows an avoided crossing with a gap of order $\Delta$.
No comparable feature is visible at $\xi=3\omega_0/2$ on this scale.
\begin{figure}
\includegraphics[width=0.9\textwidth]{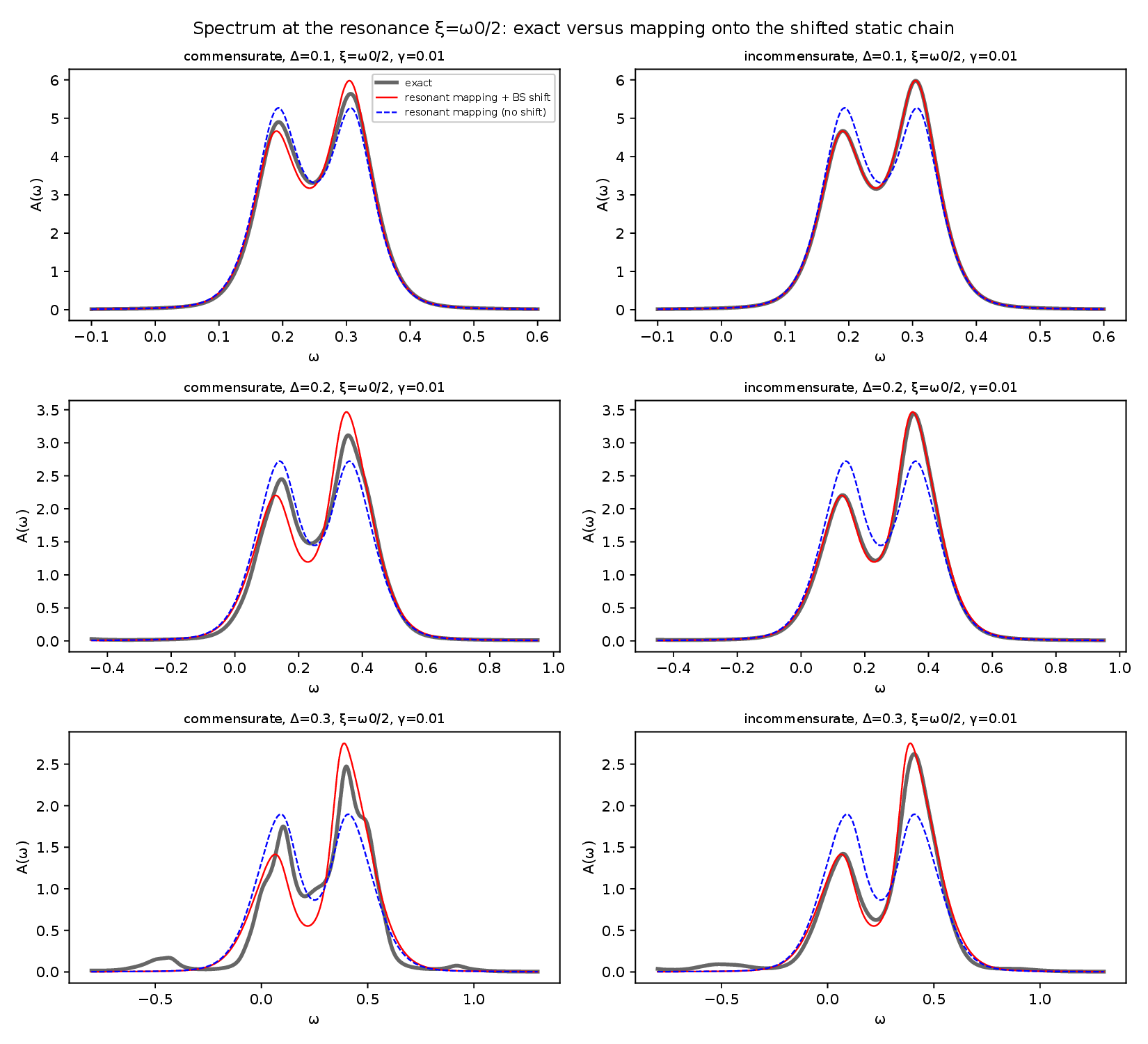}
\caption{Spectrum at $\xi=\omega_0/2$ for $\Delta=0.1,0.2,0.3$ ($\gamma=0.01$).
Grey: exact. Dashed blue: Eq.~\eqref{eq:map}. Red: Eq.~\eqref{eq:map} with the
analytic shift $\Delta^2/(4\omega_0)$. Left: commensurate,
right: incommensurate.}
\label{fig:rescuts}
\end{figure}
Figure~\ref{fig:rescuts} compares the exact spectrum at $\xi=\omega_0/2$ with
Eq.~\eqref{eq:map}. The spectrum is a doublet with a dip at its centre,
as in Eq.~\eqref{eq:Rayleigh}.
The maximum deviation relative to the peak height is:
\begin{center}\small
\begin{tabular}{c|cc|cc}
\toprule
 & \multicolumn{2}{c|}{Eq.~\eqref{eq:map}}
 & \multicolumn{2}{c}{with shift $\Delta^2/4\omega_0$}\\
$\Delta$ & comm. & incomm. & comm. & incomm.\\
\midrule
0.1 & 0.066 & 0.124 & 0.071 & 0.009\\
0.2 & 0.130 & 0.215 & 0.161 & 0.051\\
0.3 & 0.238 & 0.276 & 0.259 & 0.134\\
\bottomrule
\end{tabular}
\end{center}
The shift removes most of the deviation for incommensurate fluctuations and
does not help for commensurate ones. At $\Delta=0.1$ the exact peak positions
at $\gamma=0.01$ are $\omega=0.1905,0.3095$ (commensurate) and $0.1898,0.3052$
(incommensurate), which compares with the Rayleigh estimate
$\omega_0/2\pm\Delta/2=0.20,0.30$ (the finite $\gamma$, $\eta$ and the shifts
account for the difference).
\begin{figure*}
\includegraphics[width=0.95\textwidth]{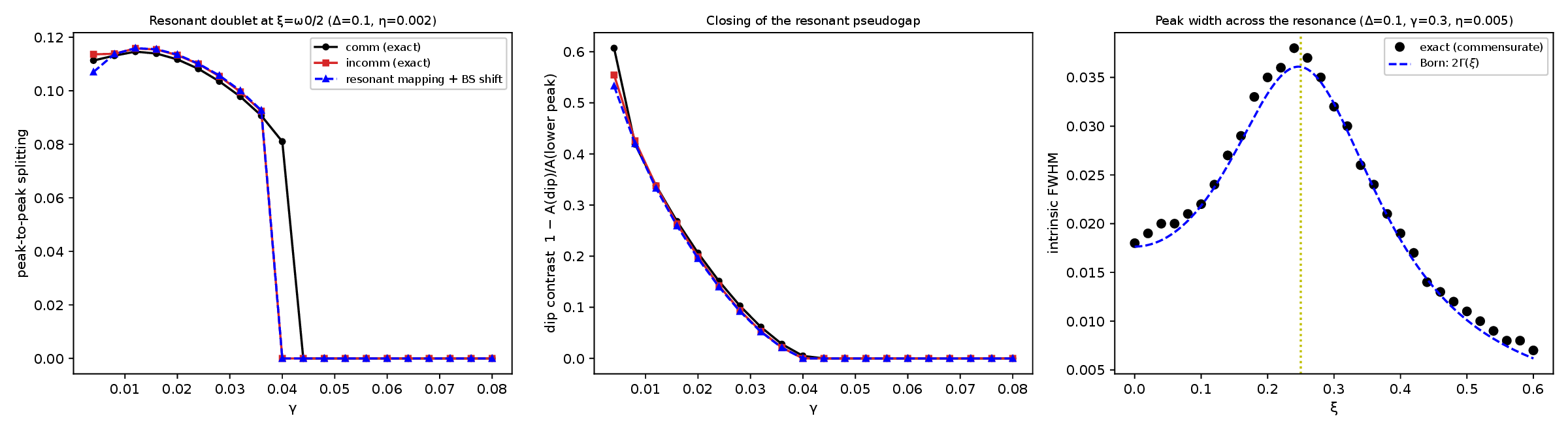}
\caption{Left: peak-to-peak splitting at $\xi=\omega_0/2$ versus
$\gamma$ ($\Delta=0.1$, $\eta=0.002$): exact (both models) and
Eq.~\eqref{eq:map} with the shift. Centre: contrast of the central dip,
$1-A_{\rm dip}/A_{\rm lower\,peak}$. Right: intrinsic FWHM of the main peak
versus $\xi$ for $\Delta=0.1$, $\gamma=0.3$, $\eta=0.005$ (commensurate),
compared with $2\Gamma(\xi)$ of Eq.~\eqref{eq:Gxi}.}
\label{fig:ressplit}
\end{figure*}
Figure~\ref{fig:ressplit} (left, centre) shows how damping affects the resonant
doublet at $\Delta=0.1$. The splitting stays at about $0.11$--$0.115$ for
$\gamma\lesssim0.02$ and decreases for larger $\gamma$; the dip contrast falls
from $0.61$ (commensurate), $0.55$ (incommensurate) and $0.53$
(Eq.~\eqref{eq:map}) at $\gamma=0.004$ to zero near $\gamma_c\simeq0.04$
(about $0.4\Delta$), where the doublet merges into a single peak.
The incommensurate model and Eq.~\eqref{eq:map} close at the same $\gamma$
(contrast $0.021$ at $\gamma=0.036$ and $0$ at $0.040$); the commensurate model
still has a small splitting ($0.081$) at $\gamma=0.040$ and closes at $0.044$.
This is the closure of the static pseudogap in the rotating frame.

For strong damping ($\gamma\gg\Delta$) the resonance appears as a peak of the
quasiparticle width. Figure~\ref{fig:ressplit} (right) shows the exact width
at $\Delta=0.1$, $\gamma=0.3$ across $\xi=\omega_0/2$: it has a maximum of
about $0.038$ at $\xi\simeq0.26$ (the second-order value is $0.036$), and
follows Eq.~\eqref{eq:Gxi} over the whole range $0\le\xi\le0.6$ to within a
few percent of the maximum. At first order in $\Delta^2$, Eq.~\eqref{eq:Gres}
gives only the resonant term of $\Gamma(\xi)$,
$\frac{\Delta^2\gamma}{2}/[(2\xi-\omega_0)^2+\gamma^2]$, so the resonant doublet
at small $\gamma$ and the Lorentzian width peak at large $\gamma$ are the same
phenomenon in two regimes.
\begin{figure*}
\includegraphics[width=0.9\textwidth]{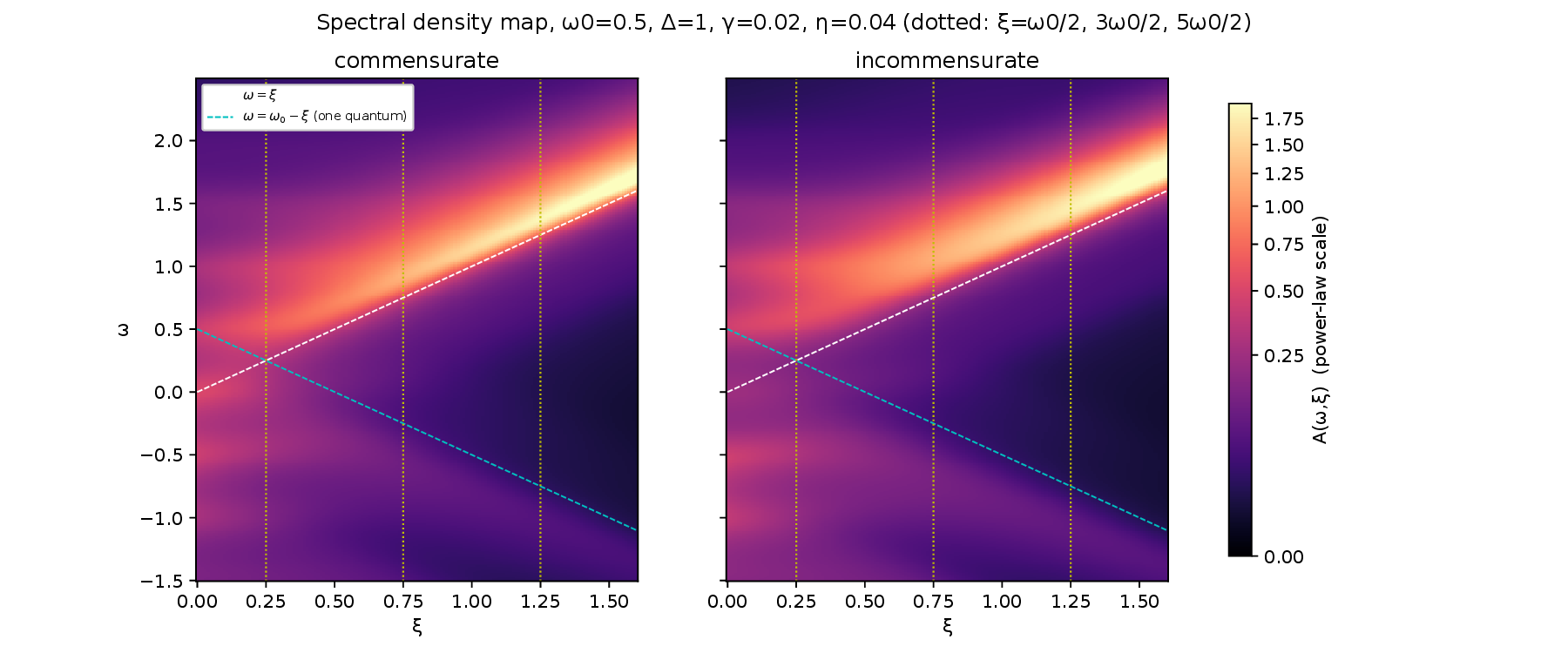}
\caption{As Fig.~\ref{fig:resmap} for $\Delta=1$, $\gamma=0.02$, $\eta=0.04$;
dotted lines at $\xi=\omega_0/2,3\omega_0/2,5\omega_0/2$.}
\label{fig:resstrong}
\end{figure*}
At the amplitude $\Delta=1$ used in the other figures (Fig.~\ref{fig:resstrong})
the resonance is no longer a distinct crossing: the main branch appears shifted
to $\omega>\xi$ for $\xi\lesssim0.6$ and the resonant doublet is merged with the
sidebands. The mapping is a weak-coupling statement ($\Delta$ small compared
with $\omega_0$): its maximum deviation grows from $1$--$7\%$ at $\Delta=0.1$
to $13$--$26\%$ at $\Delta=0.3$ in the table above, and it was not tested at
larger $\Delta$.

\subsection{Remarks}
The resonant mapping is a statement within the Fock-space formalism: it follows
from keeping the resonant block of $\hat H_{\rm eff}$, and it gives the
pseudogap near $\omega=\omega_0/2$, $\xi=\omega_0/2$ in terms of the known
static solution. Two points were not examined. First, whether the structures
at $\omega\simeq\omega_0$ and $\omega_0/2$ in the density of states of
Fig.~\ref{fig:dos} originate from the $\xi$ range near the resonance has not
been tested. Second, higher resonances $2\xi=(2m+1)\omega_0$ with $m\ge1$
involve multi-quantum processes and are not described by Eq.~\eqref{eq:map};
their effect on the spectra was not analysed.

\section{Discussion and Conclusion}
\label{sec:disc}


In the representation \eqref{eq:GR} a ``scattering history'' is a Fock 
state of the bath. A state with occupations $(n,m)$ of the modes $(a,b)$ 
has bath energy $n\varepsilon_a+m\varepsilon_b=(n-m)\omega_0-i(n+m)\gamma$, 
which is the combination that appeared as the history gap in the earlier 
double-series construction \cite{KS26}. In the exact $\hat H_{\rm eff}$ 
this energy is a scalar shift of the pseudospin energy: a transition between 
the $\xi$ and $-\xi$ branches is accompanied by creation or annihilation of 
one boson, and the intermediate energies are $\pm\xi+(\text{bath energy})$. 
This is why the exact sidebands remain close to multiples of $\omega_0$ for 
small $\xi$.


The model is classical in the field, uniform in space and Gaussian;
this is the model of Refs.~\cite{KS26,S26}. The cost of the exact solution
grows with the Poisson parameter $\lambda=\Delta^2/[2(\omega_0^2+\gamma^2)]$
\cite{KS26}, so that small $\omega_0$ and small $\gamma$ require many quanta 
and, for the incommensurate case, large matrices (dimension $\sim10^3$--$10^4$ 
in the present calculations). The use of a cutoff means that convergence must 
be checked for each parameter set.


Finally we have shown that the dynamic finite-$Q$ pseudogap model with
exponentially correlated Gaussian fluctuations of frequency $\omega_0$ and
damping $\gamma$ admits an exact solution in terms of a sparse non-Hermitian
matrix on a bosonic Fock space. The solution contains the known static continued
fractions and the commensurate Fermi-surface solution, and gives a closed 
second-order result in which the previously introduced decoherence rate 
$\Geff$ is the $\xi=0$ value of an energy-dependent rate $\Gamma(\xi)$ with 
resonances at $2|\xi|=\omega_0$. Numerically exact spectral functions and 
densities of states show sideband combs that stay close to multiples of 
$\omega_0$ at small $\xi$, a smooth non-monotonic dependence of the 
quasiparticle peak on $\gamma$, and a pseudogap that is partly filled by the 
sidebands. Near $2\xi = \omega_0$ the problem reduces to a static fuctuating-gap
problem in a rotating frame, which explains the resonant splitting of the
quasiparticle peak.

Relation of the current formalism to previously proposed \cite{KS26} approximate
Bogoliubov Ansatz is discussed in Appendix B.

\begin{acknowledgments}

The author is grateful to E.Z. Kuchinskii for many discussions 
during this work.

\end{acknowledgments}


All codes used in numerical calcuations are available from the author on 
a reasonable request.




\clearpage

\appendix

\section{Supplemental Material}

\subsection{Numerical implementation: sparse Fock-space resolvent}

The Fock-space representation provides a deterministic numerical solution of the
disorder-averaged problem.  No stochastic sampling is required: after the
exact mapping of the Gaussian fluctuating field onto a set of auxiliary
bosonic modes, the averaged retarded Green's function is a single matrix
element of the resolvent of the non-Hermitian effective Hamiltonian,
\begin{equation}
G^R(\omega,\xi)
=
\langle\psi_0|
\left(\omega-H_{\rm eff}\right)^{-1}
|\psi_0\rangle ,
\qquad
|\psi_0\rangle=|\uparrow\rangle\otimes|0\rangle .
\label{eq:fock_resolvent_numeric}
\end{equation}
The only approximation introduced in the numerical implementation is the
finite truncation of the otherwise infinite bosonic Fock space, together with
the usual infinitesimal retarded broadening used to resolve the spectral
function.

\subsubsection{Sparse linear system}

Let $\mathcal{H}_{N}$ denote the truncated Fock-space sector containing all
states retained in the numerical calculation, and let
$H_{\rm eff}^{(N)}$ be the corresponding finite matrix representation of the
effective Hamiltonian.  Rather than explicitly constructing the inverse
matrix, one solves the sparse linear system
\begin{equation}
\left[
\omega+i\eta-H_{\rm eff}^{(N)}
\right]x(\omega,\xi)
=
e_{\psi_0},
\label{eq:sparse_linear_system}
\end{equation}
where $e_{\psi_0}$ is the unit vector corresponding to the initial state
$|\psi_0\rangle$ and $\eta>0$ is a small numerical broadening.

It follows immediately that
\begin{equation}
x_{\psi_0}(\omega,\xi)
=
\langle\psi_0|
\left[
\omega+i\eta-H_{\rm eff}^{(N)}
\right]^{-1}
|\psi_0\rangle .
\label{eq:x_initial}
\end{equation}
Thus the numerical Green's function is simply
\begin{equation}
G^R_{N,\eta}(\omega,\xi)
=
x_{\psi_0}(\omega,\xi),
\label{eq:numerical_green}
\end{equation}
and the corresponding spectral function is
\begin{equation}
A_{N,\eta}(\omega,\xi)
=
-\frac{1}{\pi}
\operatorname{Im}
x_{\psi_0}(\omega,\xi).
\label{eq:numerical_spectral_function}
\end{equation}

This formulation is preferable to an explicit matrix inversion, since only
one resolvent matrix element is required.  Moreover, the effective
Hamiltonians generated by the bosonic representation are sparse: the
interaction changes the occupation of only one bosonic mode by one quantum
while simultaneously flipping the pseudospin.  The number of nonzero matrix
elements therefore grows much more slowly than the square of the dimension
of the truncated Fock space.  Sparse LU factorization can consequently be
used to solve Eq.~\eqref{eq:sparse_linear_system} independently at each
frequency.

\subsubsection{Non-Hermiticity and complex symmetry}

The effective Hamiltonian is non-Hermitian because the auxiliary bosonic
energies contain the physical decay rate,
\begin{equation}
\epsilon_j=\pm\omega_0-i\gamma .
\end{equation}
Consequently,
\begin{equation}
\operatorname{Im}H_{\rm eff}
=
-\gamma \hat N ,
\qquad
\hat N=\sum_j n_j ,
\label{eq:imag_heff}
\end{equation}
within the relevant Fock-space sector.  This negative-semidefinite imaginary
part provides the convergence of the retarded resolvent and represents the
finite correlation time of the classical fluctuating field.

At the same time, in the natural Fock basis the matrix is complex symmetric,
\begin{equation}
H_{\rm eff}^{T}=H_{\rm eff},
\qquad
H_{\rm eff}^{\dagger}\neq H_{\rm eff}.
\label{eq:complex_symmetric}
\end{equation}
The complex symmetry follows from the symmetric bosonic matrix elements,
for example
\begin{equation}
\langle n+1|a^\dagger|n\rangle
=
\sqrt{n+1}
=
\langle n|a|n+1\rangle ,
\end{equation}
whereas the complex diagonal energies prevent Hermiticity.  Thus the
calculation is naturally formulated as a non-Hermitian sparse resolvent
problem rather than as a conventional Hermitian eigenvalue problem.

\subsubsection{Fock-space truncation}

The exact Fock space is infinite.  Numerically, the bosonic occupations are
restricted to a finite range.  For the commensurate two-mode problem one may,
for example, impose
\begin{equation}
0\leq n_a\leq N_a,
\qquad
0\leq n_b\leq N_b ,
\end{equation}
while in the incommensurate four-mode problem the corresponding restrictions
are imposed on
\begin{equation}
n_{a_1},\;n_{b_1},\;n_{a_2},\;n_{b_2}.
\end{equation}
Whenever possible, the conserved parity or charge of the effective
Hamiltonian should be used first to construct only the invariant Fock-space
sector connected to $|\psi_0\rangle$.  This can reduce the numerical
dimension substantially, especially in the four-mode incommensurate case.

The required cutoff is controlled by the distribution of bosonic occupation
numbers generated by the coupling.  In particular, the natural parameter
\begin{equation}
\lambda=
\frac{\Delta^2}
     {2(\omega_0^2+\gamma^2)}
\label{eq:fock_lambda}
\end{equation}
provides a useful estimate of the characteristic occupation-number scale.
For large $\lambda$ the relevant portion of Fock space extends to larger
boson numbers, and the cutoff must be increased accordingly.  The cutoff
should therefore always be checked by explicit convergence of the calculated
spectral function rather than fixed solely from a universal prescription.

\subsubsection{Physical damping versus numerical broadening}

It is important to distinguish the physical fluctuation damping $\gamma$ from
the numerical parameter $\eta$.  The quantity $\gamma$ enters the effective
Hamiltonian itself through
\begin{equation}
\epsilon_j=\pm\omega_0-i\gamma
\end{equation}
and is therefore a physical parameter of the stochastic model.  In contrast,
$\eta$ is introduced only through
\begin{equation}
\omega\rightarrow\omega+i\eta
\end{equation}
in order to implement the retarded boundary condition and to obtain a
smooth numerical representation of the spectral density.
Thus $\gamma\neq\eta$ .
The physical spectrum should be obtained by decreasing $\eta$ until the
result is insensitive to further reduction, while keeping $\gamma$ fixed.
In the time domain, the additional $\eta$ corresponds to multiplication of
the averaged propagator by the numerical factor $e^{-\eta t}$; it does not
represent an additional physical decoherence mechanism.

\subsubsection{Convergence and numerical exactness}

There are two independent numerical limits.  The first is the removal of
the Fock-space truncation,
\begin{equation}
N\rightarrow\infty ,
\end{equation}
and the second is the retarded-resolution limit,
\begin{equation}
\eta\rightarrow0^+ .
\end{equation}
The exact result represented by the Fock-space construction is therefore
obtained as
\begin{equation}
G^R(\omega,\xi)
=
\lim_{\eta\rightarrow0^+}
\lim_{N\rightarrow\infty}
G^R_{N,\eta}(\omega,\xi).
\label{eq:double_limit}
\end{equation}

In practice, convergence should be tested by increasing the bosonic cutoff
and independently decreasing $\eta$.  For example, one can monitor
\begin{equation}
\left|
A_{N+\Delta N,\eta}(\omega,\xi)
-
A_{N,\eta}(\omega,\xi)
\right|
\end{equation}
and
\begin{equation}
\left|
A_{N,\eta/2}(\omega,\xi)
-
A_{N,\eta}(\omega,\xi)
\right|.
\end{equation}
A converged spectrum should also satisfy the spectral-weight sum rule,
\begin{equation}
\int_{-\infty}^{\infty}d\omega\,
A(\omega,\xi)=1 ,
\label{eq:spectral_sum_rule_numeric}
\end{equation}
up to the numerical accuracy associated with the finite frequency interval,
the Fock-space cutoff, and the finite value of $\eta$.

This distinction is useful for the interpretation of the term
``numerically exact.''  The mapping from the original Gaussian stochastic
problem to the Fock-space resolvent is exact and retains the full
non-commuting pseudospin dynamics.  The numerical calculation introduces
only controlled finite-Fock-space and finite-$\eta$ errors.  In particular,
there is no Monte Carlo sampling error and no Abelian or semiclassical
replacement of the time-ordered pseudospin evolution.

\subsubsection{Relation to the static continued fraction}

The continued-fraction solutions arise as special reductions of the
Fock-space representation rather than as independent approximations to it.
For example, at $\omega_0=0$ the relevant incommensurate Fock-space states
collapse onto a one-dimensional chain, and the bosonic matrix elements
generate the coefficients
\begin{equation}
v(k)=\left\lceil\frac{k}{2}\right\rceil .
\end{equation}
The corresponding resolvent therefore reduces to the familiar static
continued fraction.  In the commensurate $\omega_0=0$ case the active
oscillator instead gives
\begin{equation}
v(k)=k .
\end{equation}
At finite $\omega_0$ the Fock-space graph is higher-dimensional and the
simple one-dimensional continued fraction is generally replaced by the
full sparse resolvent.  This is precisely why the Fock-space formulation
provides a natural exact numerical treatment of the finite-frequency,
non-commuting pseudospin problem.

\begin{figure*}
\includegraphics[width=0.95\textwidth]{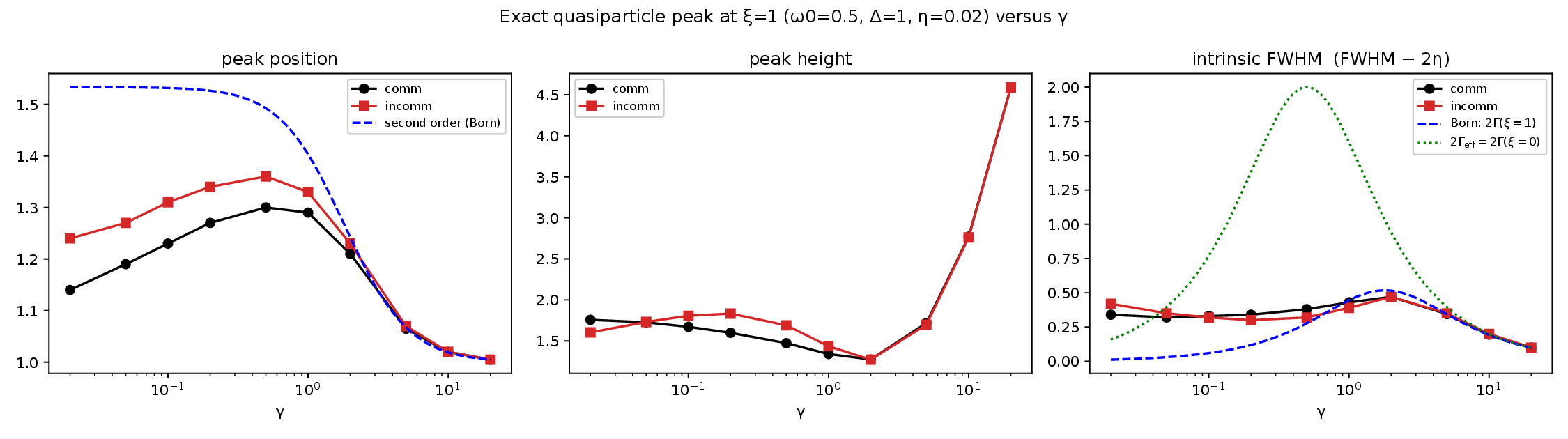}
\caption{Exact position, height and intrinsic width (FWHM minus $2\eta$) of the
main quasiparticle peak at $\xi=1$ versus $\gamma$ ($\omega_0=0.5$, $\Delta=1$,
$\eta=0.02$). Dashed: second-order (Born) formulas,
Eqs.~\eqref{eq:Sigma2}--\eqref{eq:Gxi}. Dotted: $2\Geff$.}\label{fig:peak}
\end{figure*}

\subsubsection{Matrix elements}

For the commensurate model, the off-diagonal elements of 
$\hat H^{\rm c}_{\rm eff}$ connect $\ket{\sigma;n_a,n_b}$ to 
$\ket{\bar\sigma;n_a\pm1,n_b}$ and $\ket{\bar\sigma;n_a,n_b\pm1}$ 
with amplitudes $\frac{\Delta}{\sqrt2}\sqrt{n_a+1}$, 
$\frac\Delta{\sqrt2}\sqrt{n_a}$ (and the same for $b$); 
the diagonal element is $\pm\xi+n_a\varepsilon_a+n_b\varepsilon_b$. 
For the incommensurate model, $\sigma_-$ acts on $\ket{\uparrow}$ 
with $a_i$ (lowering $n_{a_i}$) or $b_i^\dagger$ (raising $n_{b_i}$), 
and $\sigma_+$ acts on $\ket\downarrow$ with $a_i^\dagger$ or $b_i$, 
all with amplitude $\frac\Delta{\sqrt2}\sqrt{n_j+\{0,1\}}$.

\subsubsection{Derivation of the static chains}

For $\omega_0=0$ the incommensurate chain has sites $k=0,1,\dots$: 
$\ket{\uparrow;n,n}$ for $k=2n$ and $\ket{\downarrow;n,n+1}$ for $k=2n+1$. 
The hopping $\ket{\uparrow;n,n}\to\ket{\downarrow;n,n+1}$ is 
$\Delta\sqrt{n+1}$ (via $\sigma_-b^\dagger$). The hopping 
$\ket{\downarrow;n,n+1}\to\ket{\uparrow;n+1,n+1}$ is 
$\Delta\sqrt{n+1}$ (via $\sigma_+a^\dagger$). 
Hence $t_{k}^2=\Delta^2\lceil k/2\rceil$ for $k\ge1$. The continued fraction 
of a tridiagonal resolvent with diagonal $E_k=(-1)^k\xi-ik\gamma$ and these 
hoppings is the formula of Sec.~\ref{sec:exact}.

\subsubsection{Size of the space}

Truncating the commensurate space at $N_c$ total quanta gives 
$D_c=(N_c+1)(N_c+2)$ states ($D_c/2$ with the parity $\Pi$). 
In the incommensurate case, the sector $C=\frac12$ contains $\ket{\uparrow}$ 
states with $n_a=n_b=n$ [$(n+1)^2$ of them] and $\ket\downarrow$ states with 
$n_b=n_a+1=m+1$ [$(m+1)(m+2)$]. Truncating at $n,m\le n_{\max}$,
\begin{equation}
D_{\rm i}=\frac{(n_{\max}+1)(n_{\max}+2)(4n_{\max}+9)}{6}.
\end{equation}
For example $D_{\rm i}=2600$ for $n_{\max}=14$ and $6853$ for $n_{\max}=20$.

\subsubsection{Quasiparticle peak and decoherence}

Figure~\ref{fig:peak} follows the main peak at $\xi=1$ for $\gamma$ from 
$0.02$ to $20$. The exact position, height and width vary smoothly and 
non-monotonically with $\gamma$: the intrinsic width has a maximum of about 
$0.47$ near $\gamma\simeq2$ and decreases both for smaller and larger 
$\gamma$ (motional narrowing). For $\gamma\ge5$ the second-order expressions 
describe the exact peak accurately:
\begin{center}\small
\begin{tabular}{c|cc|cc}
\toprule
$\gamma$ & position (ex) & position (Born) & width (ex) & width (Born)\\
\midrule
5 & 1.065 & 1.068 & 0.345 & 0.343\\
10 & 1.020 & 1.019 & 0.195 & 0.192\\
20 & 1.005 & 1.005 & 0.100 & 0.099\\
\bottomrule
\end{tabular}
\end{center}
(commensurate fluctuations; the incommensurate values are $1.07,1.02,1.005$ 
and $0.35,0.20,0.10$). At $\gamma=2$ the Born values $1.242$ and $0.515$ are 
within $0.03$ and $0.05$ of the exact $1.21$ and $0.47$ (commensurate). 
For $\gamma\lesssim1$ the second-order result is not accurate, as expected 
for $\Delta=1$: the exact position at $\xi=1$ is $1.14$--$1.36$ whereas 
$\xi+\mathrm{Re}\,\Sigma^{(2)}\simeq1.5$. The exact width at small $\gamma$ 
is controlled by the sideband structure of the spectrum and not by a Lorentzian 
width. In the regime where the Born formulas hold, the $\xi$ dependence of 
$\Gamma(\xi)$, Eq.~\eqref{eq:Gxi}, is essential: at $\xi=1$ the exact width 
differs markedly from $2\Geff=2\Gamma(0)$ (dotted curve) for $\gamma\lesssim2$, 
because at $\xi=1$ the two Lorentzians in Eq.~\eqref{eq:Gxi} are centred at 
$2\xi=\pm\omega_0$ away from the quasiparticle energy, whereas at $\xi=0$ they 
coincide.

\section{Relation to approximate Bogoliubov Ansatz}



\subsection{The exact Fock-space representation}

We have seen above that  the classical Gaussian average
of the time-ordered exponential equals a Fock-space vacuum
expectation value:
\begin{equation}
  \left\langle
    T e^{-i\int_0^t \hat H(t')dt'}
  \right\rangle_{\Delta}
  =
  \langle 0| T e^{-i\int_0^t \hat H_{\rm eff}(t')dt'} |0\rangle,
  \label{eq:mapping}
\end{equation}
with the effective Hamiltonian
\begin{equation}
  \hat H_{\rm eff}
  =
  \xi\sigma_3
  +
  \epsilon_a n_a
  +
  \epsilon_b n_b
  +
  \frac{\Delta}{\sqrt 2}
  \sigma_1(a + a^\dagger + b + b^\dagger),
  \label{eq:Heff}
\end{equation}
where
\begin{equation}
  \epsilon_a = \omega_0 - i\gamma,
  \qquad
  \epsilon_b = -\omega_0 - i\gamma.
  \label{eq:epsilons}
\end{equation}
The retarded Green's function is
\begin{equation}
  G^R(\omega,\xi)
  =
  \langle \psi_0 |
  (\omega - \hat H_{\rm eff})^{-1}
  |\psi_0\rangle,
  \qquad
  \mathrm{Im}\,\omega > 0,
  \label{eq:GRet}
\end{equation}
with
$|\psi_0\rangle = |\uparrow\rangle\otimes|0\rangle$.

The Fock states that are reached from $|\psi_0\rangle$ are labelled
by pairs $(n_a, n_b)$. For a state $|n_a, n_b\rangle$ of the bath,
the bath energy is
\begin{equation}
  E_{\rm bath}(n_a, n_b)
  =
  n_a\epsilon_a + n_b\epsilon_b
  =
  (n_a - n_b)\omega_0
  -
  i(n_a + n_b)\gamma.
  \label{eq:bathEnergy}
\end{equation}

\subsection{Identification with Bogoliubov histories}

The Bogoliubov-Ansatz double series \cite{KS26} is
organized by ``scattering histories'' labelled by a pair
\((n,m)\), with complex Poisson weights
\begin{equation}
  P_{\lambda_1}(n) P_{\lambda_2}(m)
  =
  \frac{\lambda_1^n\lambda_2^m}{n!\,m!}
  e^{-\lambda_1-\lambda_2},
  \label{eq:Poisson}
\end{equation}
with
\begin{equation}
\lambda_1
=
\frac{\Delta^2}
{2(\omega_0-i\gamma)^2},
\qquad
\lambda_2
=
\frac{\Delta^2}
{2(-\omega_0-i\gamma)^2},
\label{lambi}
\end{equation}
and with a history-resolved complex gap
\begin{equation}
  \Delta_{nm}
  =
  (n-m)\omega_0
  -
  i(n+m)\gamma
  -
  i\Gamma_{\rm eff}.
  \label{eq:Deltanm}
\end{equation}

Comparing (\ref{eq:bathEnergy}) and (\ref{eq:Deltanm}), we find:
\begin{equation}
  E_{\rm bath}(n,m)
  =
  \Delta_{nm} + i\Gamma_{\rm eff}.
  \label{eq:identification}
\end{equation}
The bath energy and the history gap differ only by the constant
\(-i\Gamma_{\rm eff}\) that the Bogoliubov ansatz introduces to
enforce \(\langle \Delta_{nm}\rangle = 0\).

Furthermore, the squared amplitude of a Fock state
$|n_a, n_b\rangle$ from the vacuum, under the linear coupling of
\eqref{eq:Heff}, scales with the Poisson parameter
\begin{equation}
  |\langle n_a, n_b|\psi_0\rangle|^2
  \propto
  \frac{\lambda^{n_a+m_b}}{n_a!\,n_b!},
  \label{eq:FockAmplitude}
\end{equation}
which is exactly the Poisson weight \eqref{eq:Poisson}. The Bogoliubov
``histories'' are therefore Fock states of the auxiliary bosonic
bath, and the Poisson weights are the squared Fock amplitudes.

Each scattering history \((n,m)\) of the Bogoliubov ansatz
corresponds to a Fock state $|n_a, n_b\rangle$ of the auxiliary
bosonic bath of the exact solution, with:
(i) the weight \(P_{\lambda_1}(n)P_{\lambda_2}(m)\) equal to the
squared Fock amplitude;
(ii) the history gap $\Delta_{nm}$ equal to the bath energy
$E_{\rm bath}(n,m)$ shifted by \(-i\Gamma_{\rm eff}\).

\subsection{Intermediate energies: exact vs.\ Bogoliubov}

The two constructions use different expressions for the
intermediate quasiparticle energy in each history. In the exact
Fock-space solution, a history with bath occupation $(n_a, n_b)$
has the intermediate energy
\begin{equation}
  E_{n_a, n_b}^{(\pm)}
  =
  \pm\xi + n_a\epsilon_a + n_b\epsilon_b
  =
  \pm\xi + \Delta_{nm} + i\Gamma_{\rm eff},
  \label{eq:exactEnergy}
\end{equation}
which is linear in the fluctuation frequencies.

In the Bogoliubov ansatz, the intermediate energy is replaced by
the static Bogoliubov gap
\begin{equation}
  E_{nm}^{\rm Bog}
  =
  \sqrt{\xi^2 + \Delta_{nm}^2}.
  \label{eq:BogEnergy}
\end{equation}

The two expressions agree exactly at \(\xi = 0\), where
$\sqrt{\xi^2 + \Delta_{nm}^2}\big|_{\xi=0} = |\Delta_{nm}|$, and
in the static limit \(\omega_0 = 0\), where the bath energy reduces
to a purely imaginary number and both expressions become part of
the continued-fraction recursion. Away from these limits, they
differ.

\subsection{Sideband positions}

The sideband positions are the energies at which the intermediate
propagators have poles. For the $(1,0)$ history:

{\em Bogoliubov ansatz}: the sideband is at
\begin{equation}
  \omega_{\rm sideband}^{\rm Bog}
  =
  \pm\sqrt{\xi^2 + \omega_0^2}
  \approx
  \pm\left(\omega_0 + \frac{\xi^2}{2\omega_0}\right)
  \qquad
  (\xi \ll \omega_0),
  \label{eq:sidebandBog}
\end{equation}
a quadratic correction in $\xi/\omega_0$.

{\em Exact solution}: the sideband is at the intermediate energy
$-\xi + \omega_0$, i.e.,
\begin{equation}
  \omega_{\rm sideband}^{\rm exact}
  =
  -\xi + \omega_0,
  \label{eq:sidebandExact}
\end{equation}
a linear correction.

For \(\xi \ll \omega_0\), the two positions differ by
$-\xi - \xi^2/(2\omega_0)$, which is of order \(\xi\). The Monte
Carlo benchmark \cite{S26} confirmed that the exact
sideband is at $-\xi + \omega_0$, not at $+\sqrt{\xi^2+\omega_0^2}$.
This is a shortcoming of Bogoliubov Ansatz, which in other 
respects provides reasonable analytic picture of dynamic pseudogap 
within its limits (small enough $\gamma$ and $\xi$) which were 
already discussed in Refs. \cite{KS26,S26}.

\subsection{Summary of exact limits}

\begin{table}[h]
\centering
\small
\begin{tabular}{p{4cm}p{3.5cm}p{3.5cm}}
\toprule
{\em Limit} & {\em Bogoliubov ansatz} & {\em Exact Fock space} \\
\midrule
$\xi = 0$ (perfect nesting) &
  Exact (Keldysh) & Exact \\
$\omega_0 = 0$ (static) &   Exact (cont - fraction) & Exact \\
$\Delta \to 0$ (free electron) &   Free propagator & Free propagator \\
Finite \(\xi\), finite $\omega_0$ &   Sideband misplaced & Sideband correct \\
\bottomrule
\end{tabular}
\caption{Exact limits of the two constructions.}
\label{tab:limits}
\end{table}


\begin{thebibliography}{99}
\bibitem{Sad74a} M.V. Sadovskii, Sov. Phys. JETP \textbf{39}, 845 (1974).
\bibitem{Sad74b} M.V. Sadovskii, Sov. Phys. Solid State 
\textbf{16}, 1632 (1974).
\bibitem{Sad79} M.V. Sadovskii, Sov. Phys. JETP \textbf{50}, 989 (1979).
\bibitem{SadDiag} M.V. Sadovskii, \emph{Diagrammatics}, 2nd ed. 
(World Scientific, Singapore, 2019).
\bibitem{SPS98} J. Schmalian, D. Pines, and B. Stojkovi\'c, 
Phys. Rev. Lett. \textbf{80}, 3839 (1998).
\bibitem{SPS99} J. Schmalian, D. Pines, and B. Stojkovi\'c, 
Phys. Rev. B \textbf{60}, 667 (1999).
\bibitem{KS99} E.Z. Kuchinskii and M.V. Sadovskii, JETP \textbf{88}, 968 (1999).
\bibitem{KK09} M.N. Kiselev and K. Kikoin, JETP Lett. \textbf{89}, 133 (2009).
\bibitem{EK22} D.V. Efremov and M.N. Kiselev, SciPost Phys. Lect. Notes 
{\bf 65} (2022), arXiv:2206.06609.
\bibitem{KS24} E.Z. Kuchinskii and M.V. Sadovskii, 
JETP \textbf{166}, 42 (2024), arXiv:2401.08118.
\bibitem{KS26} E.Z. Kuchinskii and M.V. Sadovskii, Dynamic Pseudogap Model, 
arXiv:2608.19592.
\bibitem{S26} M.V. Sadovskii, Dynamic Pseudogap Model: Analytic Results vs 
Monte Carlo Numerics, arXiv:2610.33436.
\end{thebibliography}
\end{document}